\documentclass[twocolumn,prl,nofootinbib]{revtex4}
\usepackage{amsmath}
\usepackage{graphicx}
\usepackage{dcolumn}
\usepackage{bm}
\usepackage{amssymb}
\usepackage{latexsym}

\def\be{\begin{equation}}
\def\ee{\end{equation}}
\def\ba{\begin{eqnarray}}
\def\ea{\end{eqnarray}}

\begin{document}

\title{N-flation from multiple DBI type actions}

\author{Yi-Fu Cai$^{a}$\footnote{Email: caiyf@ihep.ac.cn} and Wei Xue$^{b,a}$\footnote{Email: wei.xue@mail.mcgill.ca}}

\affiliation{${}^a$ Institute of High Energy Physics, Chinese
Academy of Sciences, P.O.Box 918-4, Beijing 100049,
P.R.China}\affiliation{${}^b$ School of Physics, Peking
University, Beijing 100871, P.R.China}

\begin{abstract}

In this letter we present a new N-flation model constructed by
making use of multiple scalar fields which are being described by
their own DBI action. We show that the dependence of the e-folding
number and of the curvature perturbation on the number of fields
changes compared with the normal N-flation model. Our model is
also quite different from the usual DBI N-flation which is still
based on one DBI action but involves many moduli components. Some
specific examples of our model have been analyzed.

\end{abstract}

%\pacs{98.80.Cq}

\maketitle

%\section{Introduction}

Inflation \cite{inflation_bible,Starob} naturally resolves the
flatness, homogeneity and primordial monopole problems, and
predicts a scale-invariant curvature spectrum consistent with
current cosmological observations\cite{Komatsu:2008hk} very well.
So it has becomed the prevalent paradigm to understand the initial
stage of our universe. However, an inflationary model with a
single scalar generally suffers from fine tuning problems on the
parameters of its potential, such as the mass and coupling of this
field.

It was firstly noticed by Liddle {\it et al.}\cite{Liddle:1998jc}
that, when a number of scalar fields are involved, they can relax
many limits on the single scalar inflationary model. Usually,
these fields are able to work cooperatively to give a enough long
inflationary stage, even none of them can sustain inflation
separately. Models of this type have been considered later in
Refs.\cite{Malik:1998gy,Kanti:1999vt,Copeland:1999cs,Green:1999vv}.
The main results show that both the e-folding number ${\cal N}$
and the curvature perturbation $\zeta$ are approximately
proportional to the number of the scalars $N$. Later, the model of
N-flation was proposed by Dimopoulos {\it et
al.}\cite{Dimopoulos:2005ac}, which showed that a number of axions
predicted by string theory can give rise to a radiatively stable
inflation. This model has explored the possibility for an
attractive embedding of multi-field inflation in string theory.

Over the past several years, based on the recent developments in
string theory, there have been many cosmological studies on its
applications to the early universe, especially to inflation.
However, people still often encounter fine tuning and
inconsistency problems when they try to combine string theory with
cosmology. For example, Kofman and Linde in
Ref.\cite{Kofman:2002rh} pointed out the deficiency of tachyon
inflation; and there exists an $\eta$-problem in slow-roll brane
inflation as reviewed in Ref.\cite{Cline:2006hu}; and so on.
Facing to these embarrassments, it is usually suggested that
N-flation is able to relax these troubles and so can let stringy
cosmology survive. A good example is that Piao {\it et al.} have
successfully applied assisted inflation mechanism to amend the
problems of tachyon inflation\cite{Piao:2002vf}. There are also
many other works on investigating multi-field inflation models in
stringy cosmology, for example see Refs.
\cite{Majumdar:2003kd,Brandenberger:2003zk,Becker:2005sg,Cline:2005ty}.

Recently, an interesting inflationary model, which has a
non-canonical kinetic term inspired by string theory, was studied
intensively in the literature. This model is described by a
Dirac-Born-Infeld-like (DBI)
action\cite{Aharony:1999ti,Myers:1999ps}. The inflation model with
a single DBI field was investigated in
detail\cite{Silverstein:2003hf,Chen:2005ad}, which has explored a
window of inflation models without flat potentials. In this model,
a warping factor was applied to provide a speed limit which keeps
the inflaton near the top of a potential even if the potential is
steep.

In this letter, we study a multi-field inflationary model, where
each field is described by a DBI action and the total action is
constructed by the sum of them. Therefore, it is worth emphasizing
that our model is different from the usual DBI N-flation in which
only multiple moduli fields are involved in one DBI
action\cite{Huang:2007hh,Langlois:2008wt,Contaldi:2008hr}, but
ours is constructed by multiple DBI type actions (``DBIs"). This
action can be achieved if we consider a number of D3-branes in a
background metric field with negligible covariant derivatives of
field strengths and we assume that these branes are decoupled from
others. Besides, we also need to neglect the backreaction of those
branes on the background geometry as is usually done in brane
inflation models. In this scenario, the scalars are able to work
cooperatively like those in usual N-flation models. However, since
their kinetic terms are of non-canonical form, the cumulative
effect from multiple fields does not grow in linear form. From our
analysis, the e-folding number ${\cal N}$ is no longer
proportional to $N$ but to $\sqrt{N}$ instead, and the curvature
perturbation $\zeta$ is approximately proportional to $N^{3/2}$.
Thus N-flation of this type shows quite different features from
those in the usual N-flation model.

%\section{The model}

Our model is given by the following action
\begin{eqnarray}
S=\int d^4x\sqrt{-g} \bigg[ \sum_I P_I(X_I,\phi_I) \bigg]~,
\end{eqnarray}
which involves $N$ scalar fields, with
\begin{eqnarray}
P_I(X_I,\phi_I)=\frac{1}{f(\phi_I)}[1-\sqrt{1-2f(\phi_I)X_I}]-V_I(\phi_I)~,
\end{eqnarray}
where we define
$X_I\equiv-\frac{1}{2}g^{\mu\nu}\partial_\mu\phi_I\partial_\nu\phi_I$
and the sign of metric is adopted as $(-,+,+,+)$ in this letter.
This model involves multiple DBI type actions which give
the effective description of D-brane dynamics (for example see
Refs. \cite{Myers:1999ps, Maldacena:1997re}). Considering a system
constructed by a number of D3-branes in a background metric field
with negligible covariant derivatives of the field strengths and
assuming that these branes are decoupled from each other, this system
could be described by the above action which has a stringy origin
as shown in Ref.\cite{Taylor:1999pr} .

Here the scalar $\phi_I$ is interpreted as the position of the
$I$-th brane, and the warping factor
$f(\phi_I)=\frac{\lambda}{\phi_I^4}$ is suitable for all scalars
when we take on AdS-like throat and neglect the backreaction of
the branes upon the background geometry. This assumption can be
satisfied when the contribution of the background flux is much
larger than that from the branes.

We now define a series of useful parameters, (i.e. the sound
speeds), for the scalars
\begin{eqnarray}
c_{sI}\equiv\sqrt{1-2f(\phi_I)X_I}~,\label{cs}
\end{eqnarray}
which lead to interesting features of the model. We asuume spatial homogenity and isotropy, i.e. take
a flat Friedmann-Robertson-Walker metric ansatz
$ds^2=-dt^2+a^2dx^idx^i$, where $a(t)$ is the scale factor of the
universe. Then %$\dot\phi_I$ can be directly derived from
Eq.(\ref{cs}) yields %we can obtain
\begin{eqnarray}
|\dot\phi_I|=\phi_I^2(\frac{1-c_{sI}^2}{\lambda})^{\frac{1}{2}}~.
\end{eqnarray}
If $c_{sI}\sim 1$, this model returns to the slow-roll version.
However, if $c_{sI}\sim 0$, then $|\dot\phi_I|
\simeq\phi_I^2/\sqrt{\lambda}$, and in this case there is an
interesting relation for all the scalars:
\begin{eqnarray}\label{deltat}
\Delta \phi_I^{-1}=\frac{\Delta t}{\sqrt{\lambda}}~,
\end{eqnarray}
which means for a fixed time interval $\Delta t$, the variations
of $\phi_I^{-1}$ for all the scalar fields are the same.

By varying with respect to the scalar, we obtain the equations of motion:
\begin{eqnarray}
\ddot\phi_I+3H\dot\phi_I-\frac{\dot
c_{sI}}{c_{sI}}\dot\phi_I-c_{sI}P_{I,I}=0~,
\end{eqnarray}
where ``," denotes the derivative with respect to the scalar
$\phi_I$, and $H$ is the Hubble parameter defined as $\dot a/a$.

As an example, we focus on the case of IR type
potential\footnote{see Refs. \cite{Chen:2005ad} for detailed
analysis on this type DBI models.}
\begin{eqnarray}
V_I= V_{0I}-\frac{1}{2} m_I^2 \phi_I^2 ~.
\end{eqnarray}
The first part of the potential $V_{0I}$ origins from the
anti-brane tension from other throat. In IR DBI inflation,
D-branes roll from the tip of the brane, thus the potential
contains the terms like tachyon. We will assume
$s_I\equiv\frac{\dot c_{sI}}{Hc_{sI}}$ to be small numbers for
simplicity, and take the normalization $8\pi G=1$. Due to the
warping factor $f(\phi_I)$, those scalars are able to stay near
the top of their potentials, and so we have $H^2\simeq
\frac{1}{3}\sum_I V_{0I}$.

%\subsection{The efolding number}

Applying the relation (\ref{deltat}), the e-folding number of this
multiple field inflation model can be evaluated as follows,
\begin{eqnarray}\label{efold}
{\cal N}&\equiv&\int_i^f H dt \simeq
\sqrt{\frac{\lambda}{3}\sum_I V_{0I}} ~ \langle\frac{1}{\phi^i}-\frac{1}{\phi^f}\rangle \nonumber\\
&\simeq& \sqrt{N}\sqrt{\frac{\lambda}{3} \langle V_{0} \rangle} ~
\langle\frac{1}{\phi^i}\rangle ~,
\end{eqnarray}
under the assumption $c_{sI}\sim0$. Here we define $\langle {\cal
O} \rangle = (\sum_I{\cal O}_I)/N$ which is the average value of
the variables ${\cal O}_I$, and the subscript ``i" and ``f"
represent the initial and final state respectively. Since in IR
type models the scalars start rolling on the top of their
potentials\footnote{The initial condition of inflation is
essential, which is analyzed in
\cite{Goldwirth:1991rj,Brandenberger:2003py}.}, we have
$\phi^i\ll\phi^f$, and we can neglect the contribution of $\phi^f$
in Eq.(\ref{efold}). Furthermore, from Eq.(\ref{efold}) we can
deduce that the e-folding number in multiple DBIs model is
proportional to the square root of the number of scalars, i.e.
\begin{eqnarray}
{\cal N}\propto\sqrt{N}~.
\end{eqnarray}
This result is completely different from that obtained in
slow-roll N-flation which gives ${\cal N}\propto N$. This
difference shows that, in the inflationary model constructed by
multiple DBI terms, although the fields work cooperatively, the
cumulative effect from multiple fields does not grow linearly.
This results in a lot of interesting phenomena.

%\subsection{Curvature perturbation}

We now investigate the curvature perturbation of N-flation
constructed by multiple DBIs. In the calculation, we use the
Sasaki-Stewart formulism \cite{Sasaki:1995aw}, in which the
curvature perturbation on comoving slices can be expressed as the
fluctuation of the e-folding number and thus can be given in terms of
fluctuations of scalar fields $\delta\phi_I=\frac{H}{2\pi}$ on
flat slices after horizon crossing. It is given by
\begin{eqnarray}\label{zeta}
P_\zeta^{\frac{1}{2}}&=&\sqrt{\sum_{IJ}{\cal N}_{,I}{\cal
N}_{,J}<|\delta\phi_I\delta\phi_J|>}\nonumber\\
&\simeq& N^{\frac{3}{2}} \frac{\sqrt{\lambda}}{6\pi} \langle
V_{0}\rangle \sqrt{\langle\phi^{-4}\rangle} ~,
\end{eqnarray}
where we have applied the general relation ${\cal
N}_{,I}=\frac{H}{\dot\phi_I}$. This result is consistent with
single DBI inflation model when $N=1$, but if one introduces more
fields, $P_\zeta^{1/2}$ grows proportional to $N^{\frac{3}{2}}$
which is more rapid than that obtained in normal N-flation (for
example see Refs. \cite{Dimopoulos:2005ac,Kim:2006ys} and
references therein). From Eqs. (\ref{efold}) and (\ref{zeta}), we
can establish the relation between the curvature perturbation and
the e-folding number as follows,
\begin{eqnarray}\label{pzeta}
P_\zeta=\frac{{\cal N}^4
N}{4\pi^2\lambda}\frac{\langle\phi^{-4}\rangle}{\langle\phi^{-1}\rangle^4}~.
\end{eqnarray}

%\subsection{Spectral index}

Moreover, for a set of the above uncoupled fields, we can derive the
spectral index as follows,
\begin{eqnarray}
n_s-1 &\equiv& \frac{d\ln P_{\zeta}}{d\ln k} \nonumber\\
&\simeq&-2\epsilon-\frac{\sum_I(s_I+\eta_I)/(c_{sI}\epsilon_I^2)}{\sum_J1/(c_{sJ}\epsilon_{J}^2)}~,
\end{eqnarray}
where we have defined the slow-roll parameters
$\epsilon\equiv-\frac{\dot H}{H^2}$,
$\epsilon_I\equiv\frac{\dot\phi_I}{\sqrt{2c_{sI}}H}$, and
$\eta_I\equiv 2\frac{\dot\epsilon_I}{\epsilon_IH}$. When there is
only one scalar field, the above spectral index returns to the
standard form of single DBI model \cite{Chen:2006nt}. Note that
there is a relation
$\epsilon=\sum_I\epsilon_I^2\simeq\sum_I\frac{3\phi_I^4}{2c_{sI}\lambda}
/ \sum_J V_{0I}$, and this quantity can be very small when
$\lambda$ is taken to be sufficiently large. And if $\epsilon\ll
1$, each positive component $\epsilon_I$ becomes negligible
automatically. Explicitly, for the case of IR type potential we
considered currently, the spectral index can be given by
\begin{eqnarray}
n_s-1\simeq -\frac{4}{{\cal N}}\frac{\langle\phi^{-1}\rangle
\langle\phi^{-3}\rangle}{\langle\phi^{-4}\rangle}~.
\end{eqnarray}
Although it is hard to judge in general whether the spectral index of our
model is redder or bluer than that of its corresponding single
scalar model, we may study their question in certain cases. For example, the
spectral index coincides with that of the corresponding single
field model when all the scalars at the horizon-crossing time have the
same value $\phi_I=\phi_0$.

Now let us consider some specific examples of this model. The
simplest case is to choose all the scalars to have the
same value: %at the horizon-crossing time:
$\phi_I=\phi_0~$ for $I=1,...,N$. Therefore we obtain
\begin{eqnarray}\label{casea}
P_\zeta=\frac{{\cal N}^4 N}{4\pi^2\lambda}~,~~n_s=1-\frac{4}{{\cal
N}}~.
\end{eqnarray}
As is known, we need the e-folding number for inflation ${\cal
N}\simeq 60$ to explain the flatness of our universe. From the
above equation, one can easily obtain a scale-invariant spectrum
with an amplitude of order ${\cal O}(10^{-9})$ (required by
cosmological observations, such as Ref.\cite{Komatsu:2008hk}) if
$\lambda/N\sim10^{14}$.

\begin{figure}[htbp]
\includegraphics[scale=0.8]{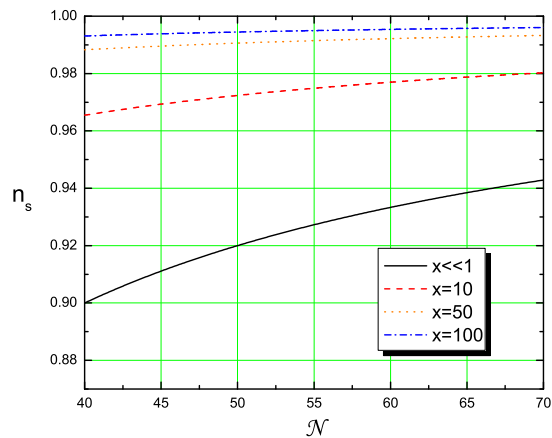}
\caption{$n_s$ as the function of the e-folding number ${\cal N}$
for different values of the variable $x$($\equiv
N\cdot\Delta/\phi_0$). The black solid line denotes the spectral
index in the first case when all the scalars have the same value
at horizon-crossing; the red dashed line denotes the spectral
index in the second case with $x=10$; the orange dotted line
$x=50$; the blue dash-dotted line $x=100$.} \label{fig:ns}
\end{figure}

Another interesting example is %that, we take the scalars as
$~\phi_I=\phi_0+I\cdot \Delta~$ for $I=1,...,N$. In order to make
this case quite different from the first one, we assume
$\phi_0\gg\Delta$ but $N\cdot\Delta\gg\phi_0$. To solve this
system, we need to apply the useful expression
\begin{eqnarray}
\langle\phi^{-l}\rangle=(-)^{l}\frac{\psi^{(l-1)}(1+\frac{\phi_0}{\Delta})-\psi^{(l-1)}(1+\frac{\phi_0}{\Delta}+N)}{(l-1)!\Delta^{l}N},
\end{eqnarray}
where $\psi^{l}(z)$ is the $l$-th derivative of the digamma
function $\psi(z)\equiv\Gamma'(z)/\Gamma(z)$. We can use the
Stirling formula to simplify the digamma function as
$\psi(z)\simeq\ln z-\frac{1}{2z}$ when $z$ is large enough.
Accordingly, we obtain the results
\begin{eqnarray}\label{caseb}
P_{\zeta}\simeq \frac{{\cal N}^4 N}{4\pi^2\lambda}
\frac{x^3}{3(\ln x)^4}~,~~n_s\simeq 1-\frac{6}{{\cal N}}\frac{\ln
x}{x}~,
\end{eqnarray}
with $x\equiv(N\cdot\Delta)/\phi_0$ in this case. From
Eq.(\ref{caseb}), for a given the e-folding number ${\cal N}$, one
can find that the tilt of the spectral index in multiple DBIs
model is strongly suppressed by the variable $x$. The dependence
of $n_s$ on the e-folding number ${\cal N}$ for different value of
the variable $x$ is plotted in Fig.\ref{fig:ns}. From the figure,
we can see that the spectrum of multiple DBIs model is generally
closer to be scale-invariant when $x$ is larger.

%\section{Conclusion and discussions}

The inflation model with multiple fields avoids some difficulties
of single field inflation models, and so is regarded as an
attractive implementation of inflation. In recent years, there
have been a number of workd studying this, such as Refs.
\cite{Piao:2006nm,Olsson:2007he,Choi:2007fya,Panotopoulos:2007pg,Battefeld:2008py},
and there is a good review on this field Ref.\cite{Wands:2007bd}.
In this letter, we have presented a new N-flation model in which a
collection of DBI fields drives inflation
simultaneously\footnote{The action of this model is similar to the
ones considered in Refs.\cite{Piao:2002vf,Ward:2007gs}, but with
different motivations.}. These scalars possess non-standard
kinetic terms, and so some non-linear information is involved when
we investigate the background evolution and curvature
perturbation. For example, the e-folding number of this model is
no longer proportional to the number of scalars, but its square
root instead as shown in Eq.(\ref{efold}). In the detailed
calculation, we considered a tachyonic potential and specifically
chose two different cases. In the first case, we took all the
scalars to have the same value at the horizon-crossing time, and
the spectral index in this case coincides with that in the single
DBI model; while in the second case, we assumed that the
collection of the scalars at the horizon-crossing time is an
arithmetical progression, and we found that the spectral index
becomes closer to $1$ if the height of this progression is much
larger than the value of the first scalar.

%\section{Non-Gaussianity}

Notice that, in this letter we merely studied the adiabatic
perturbations during inflation. However, since in a model with a
number of scalars involved, there should be entropy perturbations
generated during inflation. This is an interesting issue deserving
the future study. Here we just make some comments on this point.
Since the kinetic terms are of non-linear form, the dispersion
relations for entropy perturbations are usually modified. For
example, the sound speed will affect the time of horizon-crossing.
Therefore, if the entropy perturbations contribute to curvature
perturbation at late times, they may lead to some new features of
the primordial power spectrum. For example, a large local type
non-Gaussianity is hard to generate from the adiabatic
perturbations in DBI inflation, because the sound speeds decouple
from local type non-Gaussianity and these perturbations in this
respect become like those in standard slow-roll inflation
model\cite{Chen:2006nt}. However, entropy perturbations with small
sound speeds might be different. A good example is a model of
DBI-curvaton proposed in Ref.\cite{Li:2008fm}, where a sizable
local non-Gaussianity was generated, and an even larger
non-Gaussianity might be obtained in the model of multiple
DBI-curvaton. A more detailed study is in preparation.

\textbf{Acknowledgments} We would like to thank Robert
Brandenberger, Bin Chen, Yun-Song Piao, Yi Wang and Xinmin Zhang
for useful discussions and valuable comments. This work is
supported in part by National Natural Science Foundation of China
under Grant Nos. 10533010 and 10675136 and by the Chinese Academy
of Science under Grant No. KJCX3-SYW-N2.


\begin{thebibliography}{99}

\bibitem{inflation_bible}
  A.~H.~Guth,
  %``The Inflationary Universe: A Possible Solution To The Horizon And Flatness
  %Problems,''
  Phys.\ Rev.\  D {\bf 23}, 347 (1981);
  A.~D.~Linde,
  %``A New Inflationary Universe Scenario: A Possible Solution Of The Horizon,
  %Flatness, Homogeneity, Isotropy And Primordial Monopole Problems,''
  Phys.\ Lett.\  B {\bf 108}, 389 (1982);
  A.~Albrecht and P.~J.~Steinhardt,
  %``Cosmology For Grand Unified Theories With Radiatively Induced Symmetry
  %Breaking,''
  Phys.\ Rev.\ Lett.\  {\bf 48}, 1220 (1982).

\bibitem{Starob}
  For some early attempts we refer to:
  A.~A.~Starobinsky,
  %``A new type of isotropic cosmological models without singularity,''
  Phys.\ Lett.\  B {\bf 91}, 99 (1980);
  K.~Sato,
  %``First Order Phase Transition Of A Vacuum And Expansion Of The Universe,''
  Mon.\ Not.\ Roy.\ Astron.\ Soc.\  {\bf 195}, 467 (1981).

\bibitem{Komatsu:2008hk}
  E.~Komatsu {\it et al.}  [WMAP Collaboration],
  %``Five-Year Wilkinson Microwave Anisotropy Probe (WMAP\altaffilmark 1 )
  %Observations:Cosmological Interpretation,''
  arXiv:0803.0547 [astro-ph].

\bibitem{Liddle:1998jc}
  A.~R.~Liddle, A.~Mazumdar and F.~E.~Schunck,
  %``Assisted inflation,''
  Phys.\ Rev.\  D {\bf 58}, 061301 (1998).
%  [arXiv:astro-ph/9804177].

\bibitem{Malik:1998gy}
  K.~A.~Malik and D.~Wands,
  %``Dynamics of assisted inflation,''
  Phys.\ Rev.\  D {\bf 59}, 123501 (1999).
%  [arXiv:astro-ph/9812204].
\bibitem{Kanti:1999vt}
  P.~Kanti and K.~A.~Olive,
  %``On the realization of assisted inflation,''
  Phys.\ Rev.\  D {\bf 60}, 043502 (1999).
%  [arXiv:hep-ph/9903524].
\bibitem{Copeland:1999cs}
  E.~J.~Copeland, A.~Mazumdar and N.~J.~Nunes,
  %``Generalized assisted inflation,''
  Phys.\ Rev.\  D {\bf 60}, 083506 (1999).
%  [arXiv:astro-ph/9904309].
\bibitem{Green:1999vv}
  A.~M.~Green and J.~E.~Lidsey,
  %``Assisted dynamics of multi-scalar field cosmologies,''
  Phys.\ Rev.\  D {\bf 61}, 067301 (2000).
%  [arXiv:astro-ph/9907223].

\bibitem{Dimopoulos:2005ac}
  S.~Dimopoulos, S.~Kachru, J.~McGreevy and J.~G.~Wacker,
  %``N-flation,''
  JCAP {\bf 0808}, 003 (2008).
%  [arXiv:hep-th/0507205].

\bibitem{Kofman:2002rh}
  L.~Kofman and A.~Linde,
  %``Problems with tachyon inflation,''
  JHEP {\bf 0207}, 004 (2002).
%  [arXiv:hep-th/0205121].
\bibitem{Cline:2006hu}
  J.~M.~Cline,
  %``String cosmology,''
  arXiv:hep-th/0612129.

\bibitem{Piao:2002vf}
  Y.~S.~Piao, R.~G.~Cai, X.~M.~Zhang and Y.~Z.~Zhang,
  %``Assisted tachyonic inflation,''
  Phys.\ Rev.\  D {\bf 66}, 121301 (2002).
%  [arXiv:hep-ph/0207143].
\bibitem{Majumdar:2003kd}
  M.~Majumdar and A.~C.~Davis,
  %``Inflation from tachyon condensation, large N effects,''
  Phys.\ Rev.\  D {\bf 69}, 103504 (2004).
%  [arXiv:hep-th/0304226].
\bibitem{Brandenberger:2003zk}
  R.~Brandenberger, P.~M.~Ho and H.~C.~Kao,
  %``Large N cosmology,''
  JCAP {\bf 0411}, 011 (2004).
%  [arXiv:hep-th/0312288].
\bibitem{Becker:2005sg}
  K.~Becker, M.~Becker and A.~Krause,
  %``M-theory inflation from multi M5-brane dynamics,''
  Nucl.\ Phys.\  B {\bf 715}, 349 (2005).
%  [arXiv:hep-th/0501130].
\bibitem{Cline:2005ty}
  J.~M.~Cline and H.~Stoica,
  %``Multibrane inflation and dynamical flattening of the inflaton potential,''
  Phys.\ Rev.\  D {\bf 72}, 126004 (2005).
%  [arXiv:hep-th/0508029].

\bibitem{Aharony:1999ti}
  O.~Aharony, S.~S.~Gubser, J.~M.~Maldacena, H.~Ooguri and Y.~Oz,
  %``Large N field theories, string theory and gravity,''
  Phys.\ Rept.\  {\bf 323}, 183 (2000).
%  [arXiv:hep-th/9905111].
\bibitem{Myers:1999ps}
  R.~C.~Myers,
  %``Dielectric-branes,''
  JHEP {\bf 9912}, 022 (1999).
%  [arXiv:hep-th/9910053].

\bibitem{Silverstein:2003hf}
  E.~Silverstein and D.~Tong,
  %``Scalar speed limits and cosmology: Acceleration from D-cceleration,''
  Phys.\ Rev.\  D {\bf 70}, 103505 (2004);
%  [arXiv:hep-th/0310221];
  M.~Alishahiha, E.~Silverstein and D.~Tong,
  %``DBI in the sky,''
  Phys.\ Rev.\  D {\bf 70}, 123505 (2004).
%  [arXiv:hep-th/0404084].
\bibitem{Chen:2005ad}
  X.~Chen,
  %``Multi-throat brane inflation,''
  Phys.\ Rev.\  D {\bf 71}, 063506 (2005);
%  [arXiv:hep-th/0408084];
  X.~Chen,
  %``Inflation from warped space,''
  JHEP {\bf 0508}, 045 (2005).
%  [arXiv:hep-th/0501184].

\bibitem{Huang:2007hh}
  M.~X.~Huang, G.~Shiu and B.~Underwood,
  %``Multifield DBI Inflation and Non-Gaussianities,''
  Phys.\ Rev.\  D {\bf 77}, 023511 (2008).
%  [arXiv:0709.3299 [hep-th]].
\bibitem{Langlois:2008wt}
  D.~Langlois, S.~Renaux-Petel, D.~A.~Steer and T.~Tanaka,
  %``Primordial fluctuations and non-Gaussianities in multi-field DBI
  %inflation,''
  Phys.\ Rev.\ Lett.\  {\bf 101}, 061301 (2008).
%  [arXiv:0804.3139 [hep-th]].
\bibitem{Contaldi:2008hr}
  C.~R.~Contaldi, G.~Nicholson and H.~Stoica,
  %``Small cosmological signatures from multi-brane models,''
  arXiv:0807.2331 [hep-th].

\bibitem{Maldacena:1997re}
  J.~M.~Maldacena,
  %``The large N limit of superconformal field theories and supergravity,''
  Adv.\ Theor.\ Math.\ Phys.\  {\bf 2}, 231 (1998).
%  [Int.\ J.\ Theor.\ Phys.\  {\bf 38}, 1113 (1999)]
%  [arXiv:hep-th/9711200].
\bibitem{Taylor:1999pr}
  W.~Taylor and M.~Van Raamsdonk,
  %``Multiple Dp-branes in weak background fields,''
  Nucl.\ Phys.\  B {\bf 573}, 703 (2000).
%  [arXiv:hep-th/9910052].

\bibitem{Goldwirth:1991rj}
  D.~S.~Goldwirth and T.~Piran,
  %``Initial conditions for inflation,''
  Phys.\ Rept.\  {\bf 214}, 223 (1992).

\bibitem{Brandenberger:2003py}
  R.~Brandenberger, G.~Geshnizjani and S.~Watson,
  %``On the initial conditions for brane inflation,''
  Phys.\ Rev.\  D {\bf 67}, 123510 (2003).
%  [arXiv:hep-th/0302222].

\bibitem{Sasaki:1995aw}
  M.~Sasaki and E.~D.~Stewart,
  %``A General Analytic Formula For The Spectral Index Of The Density
  %Perturbations Produced During Inflation,''
  Prog.\ Theor.\ Phys.\  {\bf 95}, 71 (1996).
%  [arXiv:astro-ph/9507001].

\bibitem{Kim:2006ys}
  S.~A.~Kim and A.~R.~Liddle,
  %``Nflation: Multi-field inflationary dynamics and perturbations,''
  Phys.\ Rev.\  D {\bf 74}, 023513 (2006).
%  [arXiv:astro-ph/0605604].

\bibitem{Chen:2006nt}
  X.~Chen, M.~X.~Huang, S.~Kachru and G.~Shiu,
  %``Observational signatures and non-Gaussianities of general single field
  %inflation,''
  JCAP {\bf 0701}, 002 (2007).
%  [arXiv:hep-th/0605045].

\bibitem{Piao:2006nm}
  Y.~S.~Piao,
  %``On perturbation spectra of N-flation,''
  Phys.\ Rev.\  D {\bf 74}, 047302 (2006);
%  [arXiv:gr-qc/0606034];
  I.~Ahmad, Y.~S.~Piao and C.~F.~Qiao,
  %``On Number of Nflation Fields,''
  JCAP {\bf 0806}, 023 (2008);
%  [arXiv:0801.3503 [hep-th]];
%  I.~Ahmad, Y.~S.~Piao and C.~F.~Qiao,
  %``Phase Diagram for Nflation,''
  arXiv:0809.3333 [hep-th].

\bibitem{Olsson:2007he}
  M.~E.~Olsson,
  %``Inflation Assisted by Heterotic Axions,''
  JCAP {\bf 0704}, 019 (2007).
%  [arXiv:hep-th/0702109].
\bibitem{Choi:2007fya}
  K.~Y.~Choi and J.~O.~Gong,
  %``Multiple scalar particle decay and perturbation generation,''
  JCAP {\bf 0706}, 007 (2007).
%  [arXiv:0704.2939 [astro-ph]].
\bibitem{Panotopoulos:2007pg}
  G.~Panotopoulos,
  %``Assisted chaotic inflation in brane-world cosmology,''
  Phys.\ Rev.\  D {\bf 75}, 107302 (2007).
%  [arXiv:0704.3201 [hep-ph]].
\bibitem{Battefeld:2008py}
  D.~Battefeld, T.~Battefeld and A.~C.~Davis,
  %``Staggered Multi-Field Inflation,''
  arXiv:0806.1953 [hep-th];
  T.~Battefeld,
  %``Exposition to Staggered Multi-Field Inflation,''
  arXiv:0809.3242 [astro-ph].

\bibitem{Wands:2007bd}
  D.~Wands,
  %``Multiple field inflation,''
  Lect.\ Notes Phys.\  {\bf 738}, 275 (2008).
%  [arXiv:astro-ph/0702187].

\bibitem{Ward:2007gs}
  S.~Thomas and J.~Ward,
  %``IR Inflation from Multiple Branes,''
  Phys.\ Rev.\  D {\bf 76}, 023509 (2007).
%  [arXiv:hep-th/0702229].
%  J.~Ward,
  %``DBI N-flation,''
%  JHEP {\bf 0712}, 045 (2007)
%  [arXiv:0711.0760 [hep-th]].

\bibitem{Li:2008fm}
  S.~Li, Y.~F.~Cai and Y.~S.~Piao,
  %``DBI-Curvaton,''
  arXiv:0806.2363 [hep-ph].


\end{thebibliography}
\end{document}